\documentclass[twocolumn,amsmath,amssymb,aps,superscriptaddress,showpacs]{revtex4}
\usepackage{graphicx}
\usepackage{bm}
\usepackage[usenames]{color}
\begin{document}

\title{Neutron scattering study on spin correlations and fluctuations in the transition-metal-based magnetic quasicrystal Zn-Fe-Sc}

\author{Taku J Sato}
 \email{taku@issp.u-tokyo.ac.jp}
\affiliation{Neutron Science Laboratory, Institute for Solid State Physics, University of Tokyo, 106-1 Shirakata, Tokai, Ibaraki 319-1106, Japan}
\affiliation{SORST, Japan Science and Technology Agency, Kawaguchi, Saitama, Japan}

\author{Shiro Kashimoto}
\affiliation{Division of Applied Physics, Graduate School of Engineering, Hokkaido University, Sapporo 060-8628, Japan}

\author{Chihiro Masuda}
\affiliation{Division of Applied Physics, Graduate School of Engineering, Hokkaido University, Sapporo 060-8628, Japan}

\author{Takahiro Onimaru}
\affiliation{SORST, Japan Science and Technology Agency, Kawaguchi, Saitama, Japan}
\affiliation{Neutron Science Laboratory, Institute for Solid State Physics, University of Tokyo, 106-1 Shirakata, Tokai, Ibaraki 319-1106, Japan}

\author{Isao Nakanowatari}
\affiliation{Neutron Science Laboratory, Institute for Solid State Physics, University of Tokyo, 106-1 Shirakata, Tokai, Ibaraki 319-1106, Japan}

\author{Kazuki Iida}
\affiliation{Neutron Science Laboratory, Institute for Solid State Physics, University of Tokyo, 106-1 Shirakata, Tokai, Ibaraki 319-1106, Japan}

\author{Rei Morinaga}
\affiliation{Neutron Science Laboratory, Institute for Solid State Physics, University of Tokyo, 106-1 Shirakata, Tokai, Ibaraki 319-1106, Japan}

\author{Tsutomu Ishimasa}
\affiliation{Division of Applied Physics, Graduate School of Engineering, Hokkaido University, Sapporo 060-8628, Japan}
\affiliation{SORST, Japan Science and Technology Agency, Kawaguchi, Saitama, Japan}

\date{\today}

\begin{abstract}
Spin correlations and fluctuations in the $3d$-transition-metal-based icosahedral quasicrystal Zn-Fe-Sc have been investigated by neutron scattering using polycrystalline samples.
Magnetic diffuse scattering has been observed in the elastic experiment at low temperatures, indicating development of static short-range-spin correlations.
In addition, the inelastic scattering experiment detects a $Q$-independent quasielastic signal ascribed to single-site relaxational spin fluctuations.
Above the macroscopic freezing temperature  $T_{\rm f} \simeq 7$~K, the spin relaxation rate shows Arrhenius-type behavior, indicating thermally activated relaxation process.
In contrast, the relaxation rate remains finite even at the lowest temperature, suggesting a certain quantum origin for the spin fluctuations below $T_{\rm f}$.
\end{abstract}


\maketitle


\section{Introduction}
Quasicrystal is a yet mysterious form of solids, differing from periodic crystals and random glasses~\cite{she84}.
As the quasicrystal exhibits sharp Bragg reflections in a diffraction pattern, it supposedly has a translationally invariant atomic structure.
Nevertheless, its rotational symmetry, for instance the icosahedral symmetry, is incompatible with translational invariance.
These seemingly contradicting characteristics are now understood using an idea of ``quasiperiodicity'', which has hidden translational order in higher dimensional space~\cite{yam96}.

As the quasiperiodicity is a new playground for condensed matter physicists, ordering and dynamics of magnetic moments in quasiperiodic lattice have been of fundamental interest since the discovery of quasicrystals.
A number of theoretical as well as experimental studies have been carried out to date~\cite{sat05}.
Experimentally, the most extensively investigated system may be the Zn-Mg-$R$ ($R$ = rare-earth elements) quasicrystals.
The Zn-Mg-$R$ quasicrystals have well-localized magnetic moments originating from the $R^{3+}$ ions, and thus can be regarded as quasiperiodic spin systems~\cite{ishimasa04,tak01}.
They commonly exhibit spin-glass-like freezing at low temperatures in the magnetic susceptibility measurements~\cite{fis99}.
In addition, significant short-range-spin correlations have been observed by neutron diffraction~\cite{cha97,isl98,sat98a,sat00}, indicating formation of rigid spin clusters~\cite{sat05,sat06}.

Another issue on magnetic behavior of the quasicrystals is formation of localized moments, and interplay of the localized moments and conduction electrons, in the $3d$-transition-metal-based systems.
The recently discovered Zn-Fe-Sc icosahedral quasicrystal provides a novel opportunity to study this issue~\cite{mae04,kas04}.
In these previous studies, magnetic susceptibility of the Zn-Fe-Sc quasicrystal follows the Curie-Weiss (CW) law in the wide temperature range $76 < T < 300$~K.
An effective moment of Fe estimated from the CW fit is extraordinary large as $\mu_{\rm eff} \simeq 5.3\mu_{\rm B}$.
This suggests that the Zn-Fe-Sc quasicrystal is the first magnetic quasicrystal with stable $3d$ magnetic moments.
Below $T_{\rm f} \simeq 7.2$~K, an irreversibility is seen for the field-cooled and zero-field-cooled magnetizations.
Hence, the spin freezing may likely take place at $T_{\rm f}$, as commonly seen in the magnetic quasicrystals.

Compared to the $4f$ electrons of the rare-earth elements, wave functions of the $3d$ electrons have larger spatial extent with energies closer to the Fermi level.
This may possibly result in different single-site spin fluctuations as well as inter-site spin correlations in the Zn-Fe-Sc quasicrystal from those in the rare-earth-based quasicrystals.
In the present work, to reveal possible distinct behavior in the spin correlations and fluctuations, we have undertaken the first neutron-scattering investigation on the $3d$-transition-metal-based magnetic quasicrystal Zn-Fe-Sc.

\section{Experimental details}
Polycrystalline samples of the Zn$_{77}$Fe$_{7}$Sc$_{16}$ alloy were prepared by melting constituent elements.
Stoichiometric mixture of the high purity elements, Zn (99.99\%), Fe (99.99\%) and Sc (99.9\%), was put into an Al$_2$O$_3$ crucible, and sealed in a quartz tube under an Ar gas atmosphere.
The mixture was melted at $T = 1133$~K for 3~hrs, and subsequently annealed at $T = 973$~K for 3~hrs in an electric furnace.
In order to obtain homogeneous samples, the ingots were crushed into small pieces with diameters less than 2~mm, and then re-melted at $T = 1133$~K for 3~hrs with subsequent annealing at 973~K for 20hrs.
During the heat-treatments, the crucible was placed at the lowest temperature position in the furnace to avoid the evaporation loss; the loss was less than 0.5~\% of the initial weight.
The compositions of the resulting ingots were confirmed to be identical to the nominal composition by electron probe microanalysis.
Primitive icosahedral structure of the obtained samples was confirmed by the powder X-ray diffraction.
The six-dimensional lattice parameter was determined as $a_{\rm 6D} = 7.088(1)$~\AA~\cite{els86}.
Magnetization of the obtained samples was measured using a commercial superconducting quantum interference device (SQUID) magnetometer (Quantum Design).
For comparative study between the transition-metal-based and rare-earth-based magnetic quasicrystals, a polycrystalline sample of the face-centered icosahedral Zn$_{57}$Mg$_{34}$Tb$_{9}$ quasicrystal was prepared in the same manner as Ref.~\cite{sat06}.

Neutron scattering experiments were carried out using the thermal-neutron triple-axis spectrometer ISSP-GPTAS, installed at the JRR-3M research reactor (Tokai, Japan).
The polycrystalline samples were crushed into the powder form, loaded in an Al sample cell, and attached to a closed cycle refrigerator.
Both the elastic and inelastic experiments were performed using the triple-axis mode.
The pyrolytic graphite (PG) 002 reflections were used for the monochromator and analyzer, and higher harmonic neutrons were eliminated by the PG filter.
For the inelastic experiment, we employed the final-energy fixed mode with $E_{\rm f} = 13.7$~meV.
Proper combinations of collimators were selected to satisfy contradicting necessity for energy resolution and intensity.
Typical collimations were 40'-80'-80'-80'.

\section{Results and discussion}

\subsection{Magnetization measurements}

\begin{figure}
\includegraphics[scale=0.34, angle=-90]{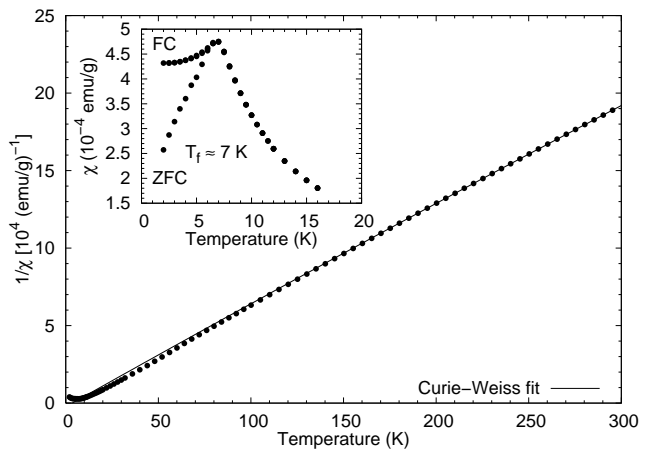}
\caption{Inverse susceptibility ($1/\chi = H/M$) obtained from the temperature dependence of the magnetization measured under the dc external field of 600~Oe.
The solid line stands for the Curie-Weiss fit in the  temperature range $200 < T < 300$~K.
Inset: magnified figure in the low temperature region ($T < 20$~K) observed under the dc field of 50~Oe. 
Field-cooled (FC) and zero-field-cooled (ZFC) susceptibilities are shown.
}
\end{figure}

The magnetization of the Zn-Fe-Sc quasicrystal has been reexamined using the presently prepared samples, since macroscopic magnetic behavior may likely depend on sample preparation condition.
Figure 1 shows the inverse susceptibility ($1/\chi = H/M$) deduced from magnetization measured under the external dc field of 600~Oe.
The inverse susceptibility shows linear behavior in a high temperature range.
Fitting the data in $200 < T < 300$~K to the CW law $\chi(T) = N_{\rm A}\mu_{\rm eff}^2/3 k_{\rm B} (T - \Theta) + \chi_0$, we obtain the optimum parameters as $\mu_{\rm eff} = 3.23(2)~\mu_{\rm B}$, $\Theta = 3(1)$~K, and $\chi_0 = 1.9(3)\times 10^{-7}$~emu/g.
The effective moment estimated here is considerably smaller than the previously reported value of $5.3~\mu_{\rm B}$~\cite{kas04}.
This may be attributable to slightly different alloy compositions of the two differently prepared samples, since formation of the Fe moment may strongly depend on small change in the Fe concentration.
The Weiss temperature is also considerably smaller than the previous value (6.54~K).
At low temperatures, there appears clear irreversibility between the field-cooled and zero-field-cooled magnetizations, suggesting spin-glass-like freezing.
The freezing temperature is determined as $T_{\rm f} \simeq 7$~K for the present sample.
Despite the significant reduction of the Weiss temperature implying weaker inter-spin interactions, the freezing temperature is in a good correspondence with the previous result.
This point will be discussed later.

\subsection{Neutron scattering formulation}
The neutron scattering function is related to the spin pair-correlations function as~\cite{lov84}:
\begin{widetext}
\begin{equation}
	S({\bm Q}, \hbar\omega)  \simeq \frac{(\gamma r_0 g)^2}{4}  \sum_{ij} 
	{\rm e}^{-{\rm i} {\bm Q} \cdot ({\bm R}_i - {\bm R}_j)} 
	\frac{1}{2\pi\hbar} \int {\rm d}t {\rm e}^{- {\rm i} \omega t}
        \langle {\bm J}_i^{\perp} \cdot {\bm J}_j^{\perp}(t) \rangle,
\end{equation}
\end{widetext}
where $\gamma = -1.91$ and $r_0$ is the electron classical radius.
${\bm J}_i^{\perp}(t)$ is defined as $\tilde{\bm Q}\times [{\bm J}_i(t) \times \tilde{\bm Q}]$, where ${\bm J}_i(t)$ is the total angular-momentum operator of the magnetic atom at the $i$-th site, and $\tilde{\bm Q}$ is the unit vector along ${\bm Q}$.
For simplicity, we assume negligible $Q$-dependence for a magnetic form factor in the above equation; this is a crude but necessary approximation, since the ionization state of the Fe atoms is unknown in the Zn-Fe-Sc system.

For the triple-axis spectrometer, the energy resolution may be given by the following Gaussian function with a full width at half maximum (FWHM) denoted by $\alpha$:
\begin{equation} 
  R(\hbar\omega) = \frac{2 \sqrt{\ln 2}}{\alpha \sqrt{\pi}} 
  \exp\left[\frac{-4 \ln 2 (\hbar\omega)^2}{\alpha^2}\right] .
\end{equation}
The neutron scattering intensity at the elastic position $I({\bm Q}, \hbar\omega^{\rm obs} = 0)$ observed in the triple-axis mode is given by a convolution of the scattering function with the energy resolution function as:
\begin{widetext}
\begin{eqnarray}
	I({\bm Q}, \hbar\omega^{\rm obs} = 0) & \simeq & 
        \int {\rm d}\hbar \omega S({\bm Q}, \hbar \omega) R(\hbar \omega) \nonumber \\ 
        & \simeq & \frac{(\gamma r_0 g)^2}{4}  \sum_{ij} 
	{\rm e}^{-{\rm i} {\bm Q} \cdot ({\bm R}_i - {\bm R}_j)} 
	\frac{1}{2\pi\hbar} \int {\rm d}t \exp\left( - \frac{\alpha^2 t^2}{16 \hbar^2 \ln 2}\right)
        \langle {\bm J}_i^{\perp} \cdot {\bm J}_j^{\perp}(t) \rangle. \label{eq:sq}
\end{eqnarray}
\end{widetext}
As seen in the above equation, the observed elastic intensity is related to the slow spin fluctuations that are within the Gaussian time window [$\exp(-4t^2\ln2/\beta^2)$] with FWHM of $\beta = 8 \hbar \ln 2/\alpha$.

At high temperatures in paramagnetic phase, the spin fluctuations become fast enough compared to $\beta$, and thus give rise to negligible scattering intensity in the elastic channel (except for the trivial self instantaneous correlations.)
Thus, one may extract the magnetic signal by taking intensity differences between the lowest and paramagnetic temperatures, assuming negligible temperature dependence in nuclear scattering.
Finally, we note that observed intensity in the present experiment corresponds to the powder spherical average as:
\begin{equation}\label{eq:powsq}
	\tilde{I}(Q, \hbar\omega^{\rm obs}) = \int \frac{{\rm d}\Omega_{\tilde{\bm Q}}}{4 \pi}  I({\bm Q}, \hbar\omega^{\rm obs}),
\end{equation}
where ${\rm d}\Omega_{\tilde{\bm Q}}$ is an element of solid angle along $\tilde{\bm Q}$.

\subsection{Neutron elastic scattering}

The neutron powder diffraction patterns have been measured at the lowest temperature $T = 4$~K and at the two paramagnetic temperatures $T = 60$~K and 200~K.
Figure~2(a) [2(b)] shows the comparison between 4~K and 60~K [200~K].
At the paramagnetic temperatures, several Bragg reflections are seen in the patterns.
Positions of the Bragg reflections are consistent with the previous X-ray results~\cite{kas03}; no extra nuclear Bragg reflections were observed.
This confirms absence of secondary-phase contamination within the present experimental accuracy.
By comparing the lowest temperature (below $T_{\rm f}$) pattern to the paramagnetic ones, we find that no magnetic Bragg peak appears at the lowest temperature.
This excludes ferromagnetic or antiferromagnetic long-range order in the Zn-Fe-Sc quasicrystal, and thus the anomaly at $T_{\rm f}$ in the susceptibility measurement is certainly due to the glassy freezing.

Intensity differences between the two patterns are also shown in the figures to increase the visibility of small temperature variation.
In the difference, the data points around the strong nuclear Bragg positions (such as at $Q \simeq 0.62$~\AA$^{-1}$) are removed, since the temperature variation of the lattice constant gives a considerable systematic error around the Bragg peak.
A salient feature of the intensity differences are appearance of a broad peak around $Q \simeq 0.6$~\AA$^{-1}$.
This clearly indicates development of the short-range-spin correlations at low temperatures.

It is worthwhile to compare the presently observed magnetic diffuse scattering with that in the rare-earth-based Zn-Mg-Tb quasicrystal.
For this purpose, the observed raw data are corrected for absorption, and scaled to the absolute units of cross section by using vanadium standard.
To compare the two systems with different magnetic elements, the resulting intensity differences $\tilde{I}^{\rm diff}(Q, \hbar\omega^{\rm obs} = 0)$ is further normalized by $(gJ)^2$:
\begin{equation}
	\tilde{I}^{\rm diff}_{\rm norm}(Q, \hbar\omega^{\rm obs} = 0) 
	= \frac{\tilde{I}^{\rm diff}(Q, \hbar\omega^{\rm obs} = 0)}{(gJ)^2}.
\end{equation}
For the Zn-Fe-Sc quasicrystal, the magnetic moment $g \mu_{B} J$ of the Fe atoms is still controversial.
We, here, use the presently determined effective moment $\mu_{\rm eff} = 3.2$~$\mu_{\rm B}$; assuming $g \sim 2$, this corresponds to $gJ \sim 2.4$.
For the Zn-Mg-Tb quasicrystal the free Tb$^{3+}$ ion value $gJ = 9$ is used.
Figure~3 shows the normalized intensity $\tilde{I}_{\rm norm}^{\rm diff}(Q, \hbar\omega^{\rm obs} = 0)$ for the Zn-Fe-Sc quasicrystal, together with that for the Zn-Mg-Tb reference.
The amplitude of the $Q$-dependence is mostly the same in both the quasicrystals.
This confirms that strength of the inter-site correlations is quantitatively comparable in the two seemingly different systems.
On the other hand, the background for the Zn-Fe-Sc quasicrystals is apparently larger than that of Zn-Mg-Tb.
This infers existence of the $Q$-independent component at low temperatures in Zn-Fe-Sc, and consequently suggests existence of slow single-site fluctuations.

\begin{figure}
\includegraphics[scale=0.34, angle=-90]{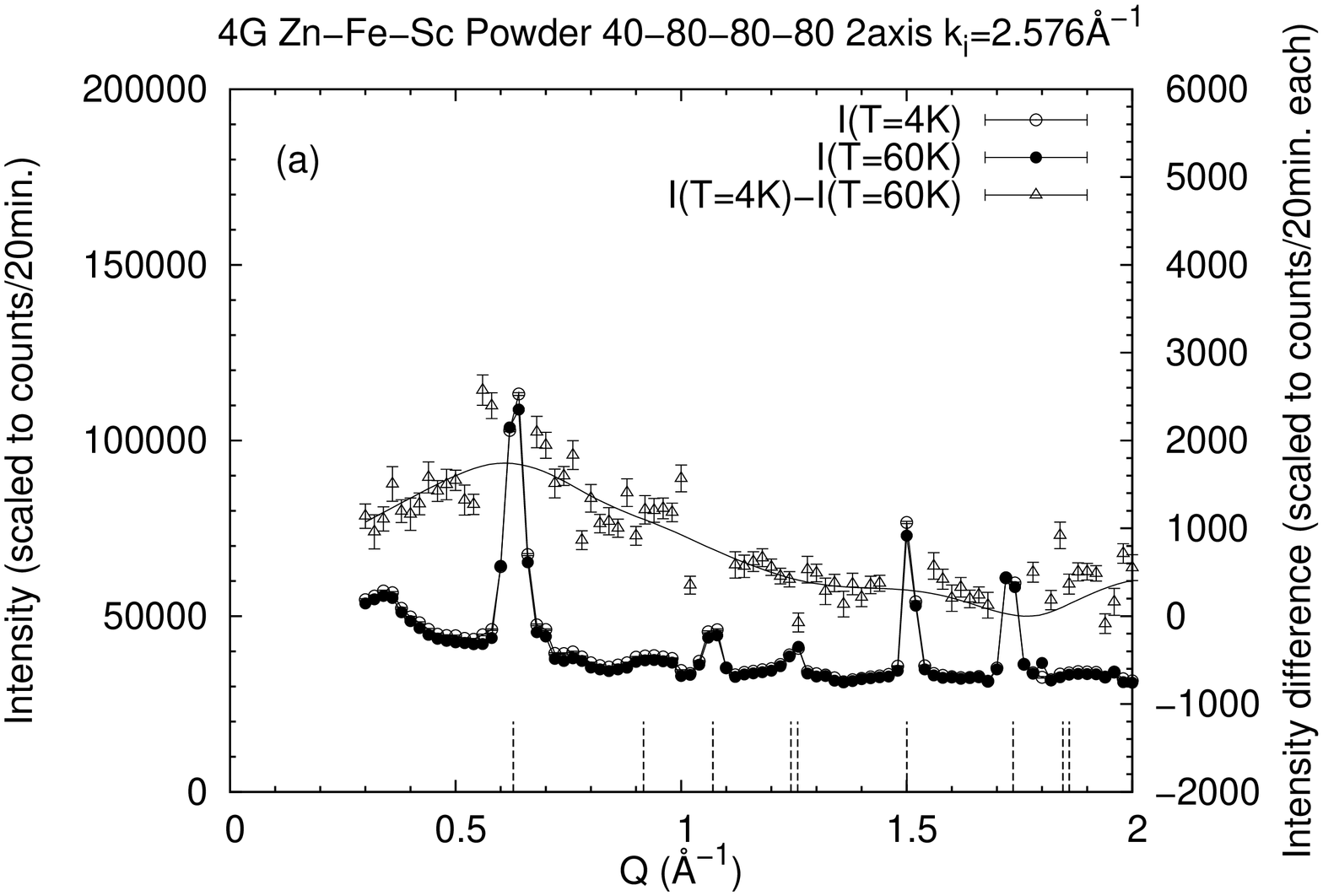}
\includegraphics[scale=0.34, angle=-90]{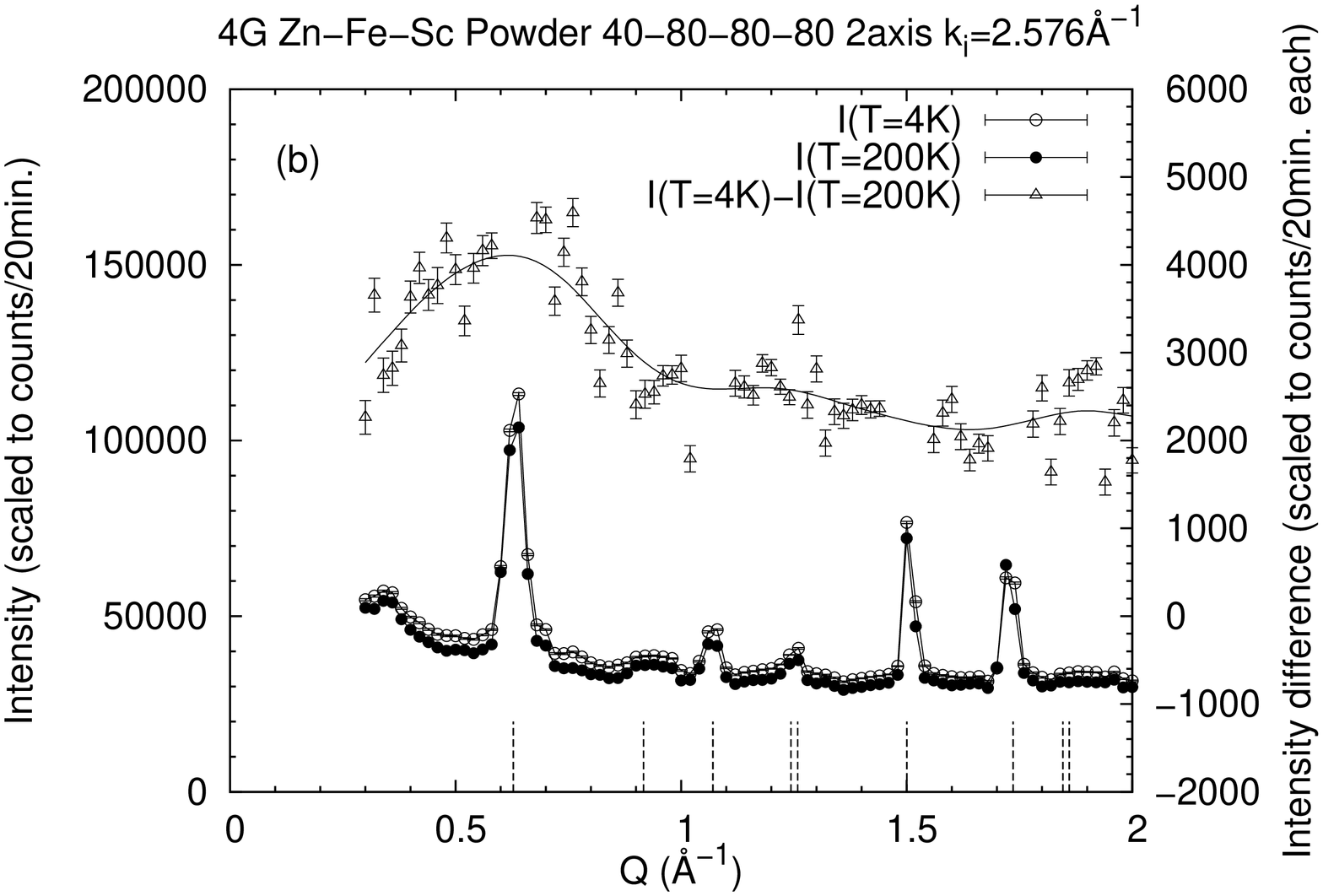}
\caption{Comparisons of powder diffraction in the Zn-Fe-Sc quasicrystal measured at (a) $T = 4$~K (open circles) and 60~K (filled circles), and (b) $T = 4$~K (open circles) and 200~K (filled circles).
Intensity differences [(a) $\tilde{I}(4~{\rm K}) - \tilde{I}(60~{\rm K})$ and (b) $\tilde{I}(4~{\rm K}) - \tilde{I}(200~{\rm K})$] are also shown by the open triangles, as an estimate of the magnetic diffuse scattering contribution.
The solid lines are guides for the eyes, whereas the dashed lines at the bottom represent nuclear Bragg reflection positions~\cite{kas03}.}
\end{figure}

\begin{figure}
\includegraphics[scale=0.32, angle=-90]{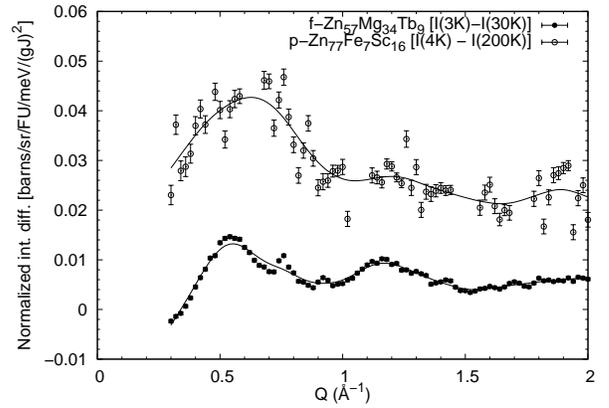}
\caption{Comparison of the normalized scattering intensity difference $\tilde{I}^{\rm diff}_{\rm norm}(Q, \hbar\omega^{\rm obs} = 0)$ obtained for the Zn-Mg-Tb and Zn-Fe-Sc quasicrystals.
The scattering intensities are for one mole of the each magnetic atom (Fe or Tb). 
See text for details.}
\end{figure}

Temperature dependence of the diffuse scattering intensity has been measured with the fixed scattering vector $Q = 0.72$~\AA$^{-1}$; this $Q$ is chosen to avoid contamination from the strong nuclear Bragg reflection at $Q \simeq 0.62$~\AA$^{-1}$.
The measurement has been carried out using the collimations 40'-80'-80'-80' with the energy resolution of $\Delta \hbar \omega = 1.14(2)$~meV at the elastic position.
Resulting temperature dependence is shown in Fig.~4(a).
The scattering intensity increases monotonically as temperature is decreased, exhibiting a weak change of slope around 200~K.
The magnetic diffuse scattering intensity in the Zn-Mg-Tb quasicrystal is also shown in the figure for comparison.
The Zn-Mg-Tb quasicrystal shows steeper increase below 50~K, which is considerably lower than the temperature at which the intensity increases in the Zn-Fe-Sc quasicrystal.
By defining the characteristic temperatures as $T_{\rm ch} = 200$~K (Zn-Fe-Sc) and 50~K (Zn-Mg-Tb), the $T_{\rm ch}$-scaled data for the two different systems shows good coincidence as shown in Fig.~4(b).
Therefore, the overall feature of the temperature dependence is quite similar; only the temperature scale is four times larger in the Zn-Fe-Sc quasicrystals.

Equation~\ref{eq:sq} indicates that the elastic intensity measured in the triple-axis mode corresponds to the time-integration of the spin-correlation function within the Gaussian time window.
It should be noted that the spin-correlation function includes the self-correlation (single-site fluctuation).
Thus, the increase of the diffuse intensity indicates not only the development of the short-range-spin correlations, but also the slowing down of the single-site spin fluctuations.
Details of the single-site fluctuation will be discussed later using the inelastic scattering technique.
We note that in both the systems the increase of the diffuse scattering intensities starts at considerably higher temperatures than the macroscopic freezing temperature $T_{\rm f} \simeq 7$~K (Zn-Fe-Sc) or $T_{\rm f} \simeq 5.8$~K (Zn-Mg-Tb).
This is due to the difference in the time scales of the macroscopic and neutron-scattering experiments~\cite{mur78}; as noted above, the neutron scattering detects the fluctuations slower than $\sim 8\hbar \ln 2/\alpha = 3.2\times10^{-12}$~s (FWHM) as elastic, whereas the time scale of the macroscopic experiments is typically of the order of 1~s.

We also note that the Weiss temperature $\Theta = 3(1)$~K is significantly smaller than the temperature scale of the short-range-spin correlations ($\sim 200$~K) in the present Zn-Fe-Sc quasicrystal.
Since the Weiss temperature is the measure of the total inter-spin interactions in the system, this discrepancy suggests that there exist both the ferro- and antiferro-magnetic interactions of similar magnitudes, canceling each other.
The cancellation may be strongly influenced by a small change in the balance of the ferro- and antiferro-magnetic interactions, possibly originating from a slight difference in composition or structural quality.
This may, at least partly, explain the amplified fluctuation of the Weiss temperatures between the samples of the present and earlier studies.
On the other hand, since the freezing temperature may be largely determined by absolute values of the inter-spin interactions, a small change in the balance may be less effective.
Although the fluctuation of the Weiss temperatures may be understood as above, that of the effective moments is not clear at the present moment.
Instead of pursuing details of the effective-moment fluctuation, here, we only note two sets of findings reported to date.
Firstly, sample-to-sample fluctuation of effective moment has been frequently reported in the transition-metal-based quasicrystals, such as the icosahedral Al-Pd-Mn quasicrystal, where even a distribution of effective moment in a single sample has been inferred~\cite{Chernikov93,Lasjaunias95,Klanjsek03}.
Secondly, recent theoretical calculation of Fe moment in the Zn-Fe-Sc approximant suggests much smaller values of $\sim 2 \mu_{\rm B}$~\cite{Ishii06}.

\begin{figure}
\includegraphics[scale=0.34, angle=-90]{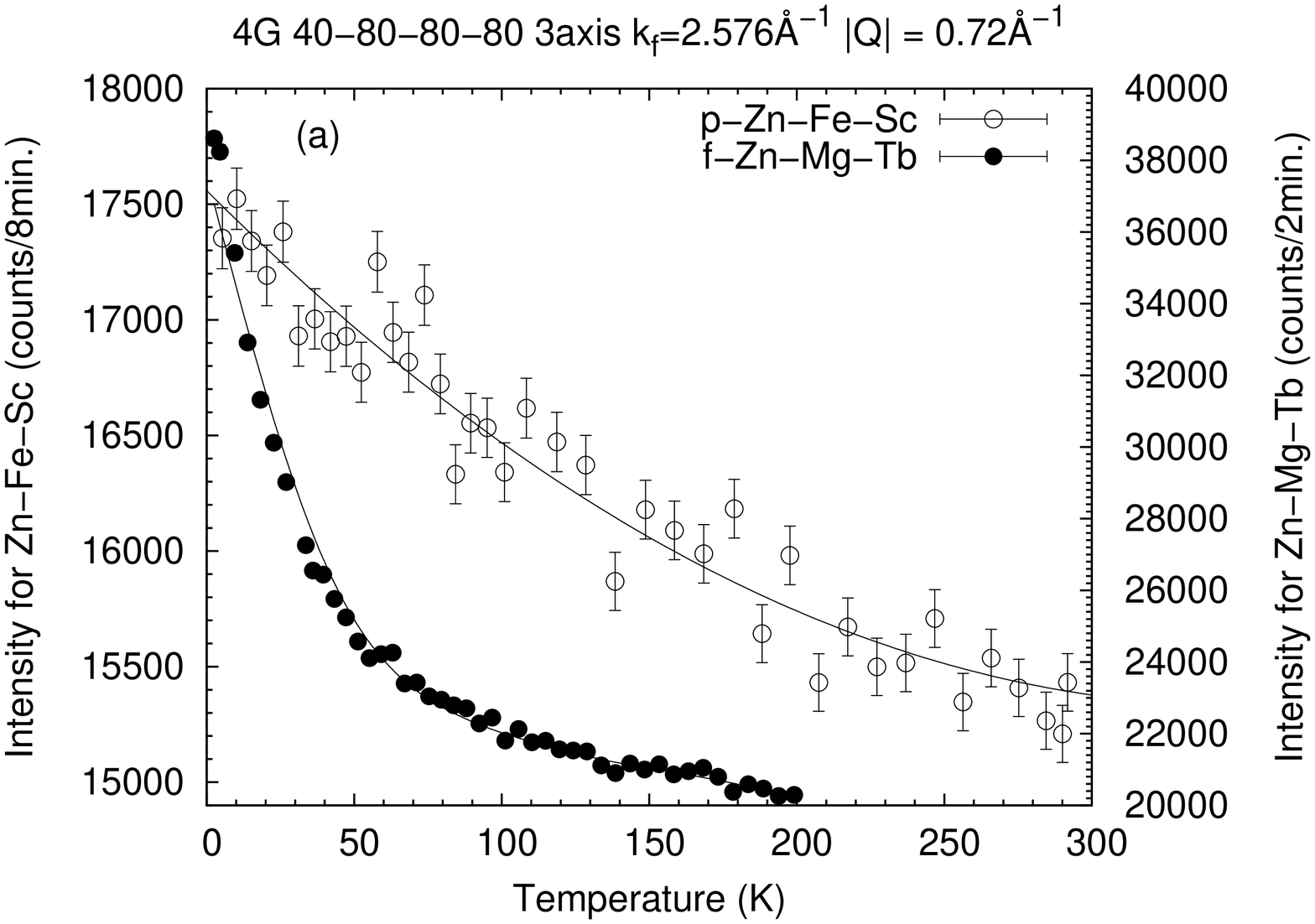}
\includegraphics[scale=0.34, angle=-90]{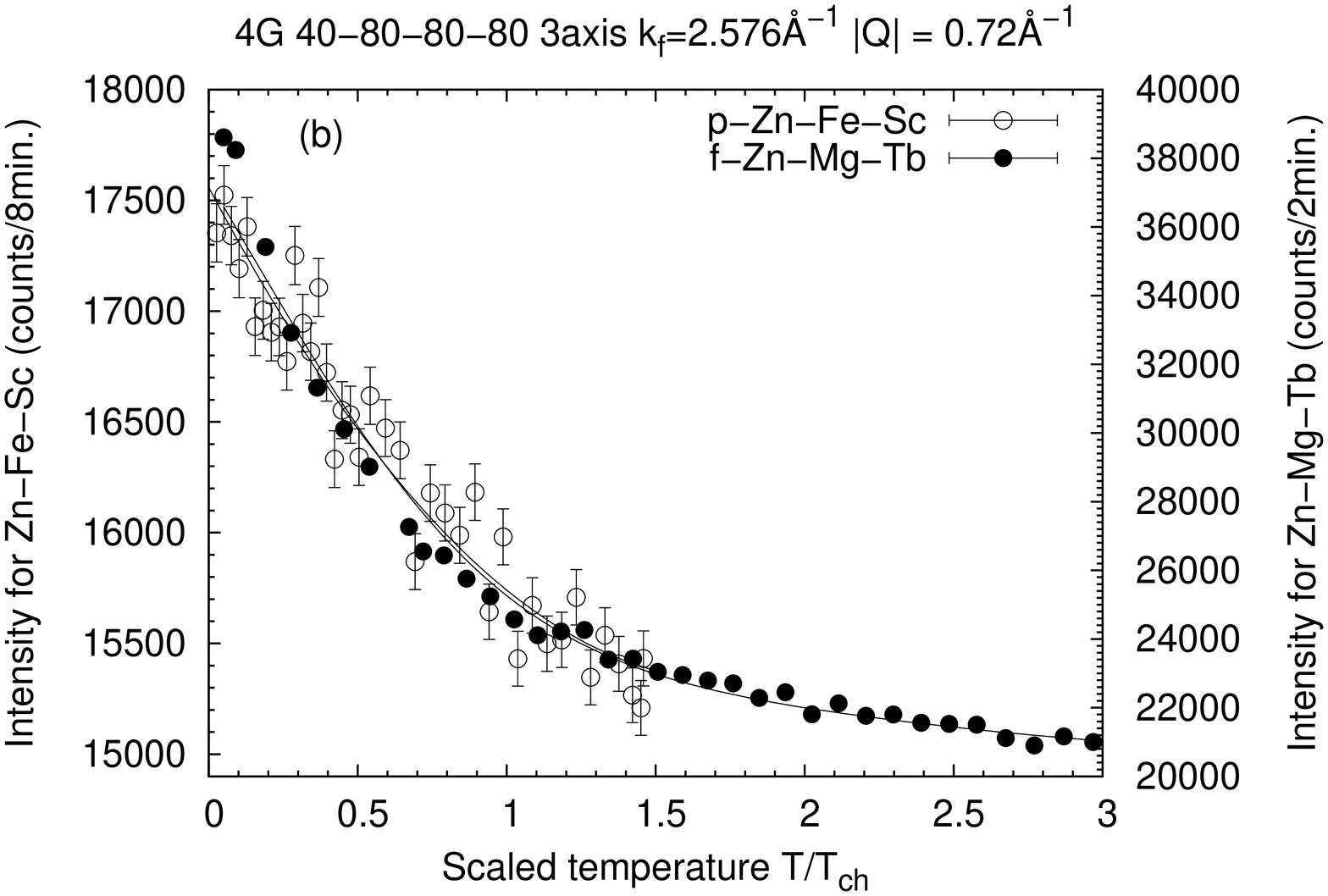}
\caption{(a) (Open circles) temperature dependence of the diffuse scattering intensity observed at $Q = 0.72$~\AA$^{-1}$ in the Zn-Fe-Sc quasicrystal.
(Closed circles) temperature dependence of the diffuse scattering intensity observed at $Q = 0.55$~\AA$^{-1}$ in the Zn-Mg-Tb quasicrystal.
The collimations were 40'-80'-80'-80', resulting in the energy resolution $\Delta \hbar \omega \simeq 1.14(2)$~meV at the elastic position.
(b) The same temperature-dependence data scaled by the characteristic temperatures $T_{\rm ch} = 200$~K (Zn-Fe-Sc) and 50~K (Zn-Mg-Tb).}
\end{figure}

\subsection{Neutron inelastic scattering}

\begin{figure}
\includegraphics[scale=0.55, angle=-90]{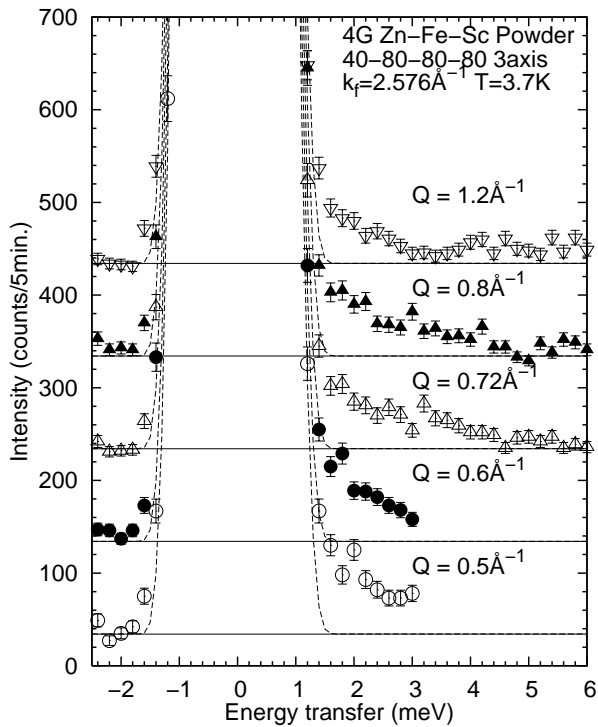}
\caption{Inelastic scattering spectra observed at $T = 3.7$~K and $Q = 0.5, 0.6, 0.72, 0.8$ and 1.2~\AA$^{-1}$ (from bottom to top).
The solid lines indicate background level, estimated from the $\hbar \omega < 0$ side, whereas the dashed lines represent an instrumental resolution function at the elastic position determined by the vanadium standard scans.
Each spectrum is shifted by 100~counts to increase the visibility.}
\end{figure}

Next, to obtain information on the spin dynamics, inelastic scattering spectra have been measured in the wide temperature range $3.7 < T < 100.4$~K.
Figure~5 shows several constant-$Q$ scans at the lowest temperature $T = 3.7$~K.
Solid lines stand for the background level estimated from the $\hbar \omega < 0$ side, whereas dashed lines denote the instrumental resolution function determined from vanadium standard scans.
It is clearly seen in the figure that significant scattering intensity exists for the $\hbar \omega > 0$ side appearing as quasielastic tails of the elastic signal.
No significant $Q$-dependence can be seen for this scattering intensity within the statistics of the present experiment; only weakly decreasing behavior for high $Q$ may be seen.
This suggests the single-site relaxational nature of corresponding spin fluctuations.
It may be noted that the weakly decreasing behavior may be due to the magnetic form factor, inferring the magnetic origin of the inelastic scattering.
The temperature dependence of the inelastic scattering has been investigated at the fixed $Q$ position $Q = 0.72$~\AA$^{-1}$; this $Q$ position is again selected so as to be close enough to the diffuse peak position ($Q \sim 0.65$~\AA$^{-1}$), and at the same time to avoid contamination from the nuclear Bragg peak situated at $Q = 0.62$~\AA$^{-1}$.
Figure~6 shows the resulting inelastic spectra observed up to $T = 100.4$~K.
The quasielastic scattering intensity increases as the temperature is elevated, and the width of the peak concomitantly grows.
It should be noted that the inelastic spectrum is totally different from the Zn-Mg-Tb quasicrystals, where inelastic scattering intensity for the $\hbar \omega > 0$ side exhibits a broad peak around $\hbar \omega \simeq 2.5$~meV, a salient feature resulting from formation of dodecahedral spin clusters~\cite{sat06}.
It also differs from the temperature-independent $S(Q, \hbar\omega)$ recently observed in the Zn-Mg-Ho quasicrystals~\cite{sato07}.
Instead, the presently observed spectrum is rather similar to energy spectra of the canonical spin glasses, where the spin dynamics is governed by relaxational process.

\begin{figure}
\includegraphics[scale=0.55, angle=-90]{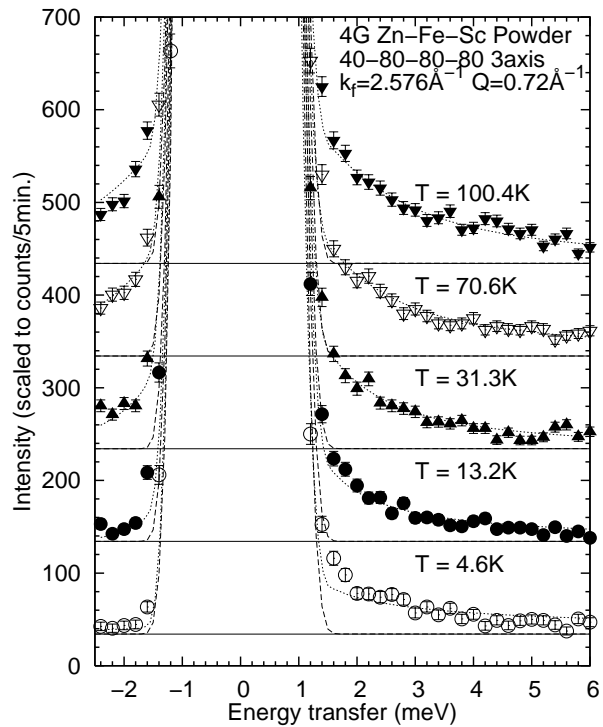}
\caption{Inelastic excitation spectra observed at $Q = 0.72$~\AA$^{-1}$ and at $T = 4.6, 13.2, 31.3, 70.6$ and 100.4~K (from bottom to top).
The solid lines indicate background level, whereas the dashed lines represent an instrumental resolution function at the elastic position.
The dotted lines are the results of resolution convoluted fitting to Eq.~(\ref{eq:lorentz}).
Each spectrum is shifted by 100~counts to increase the visibility.}
\end{figure}

To parameterize the temperature dependence of the quasielastic signal, we introduce the following trial scattering function:
\begin{eqnarray}\label{eq:lorentz}
S(Q, \hbar \omega) &=& C_{\rm qel}\frac{\Gamma_{\rm qel}}{(\hbar \omega)^2 + \Gamma_{\rm qel}^2}
\frac{\hbar\omega}{1 - e^{-\hbar\omega/k_{\rm B}T}}\nonumber\\
& + & C_{\rm el} \delta(\hbar\omega) + C_{\rm BG}.
\end{eqnarray}
The first term stands for the Lorentzian-type quasielastic component, whereas the second term is for the elastic peak.
The last term is for the constant background.
The trial scattering function is convoluted by the instrumental resolution function, and then fitted to the observed spectra.
Figure~7 shows the resulting temperature dependence of $\Gamma_{\rm qel}$ at $Q = 0.72$~\AA$^{-1}$.
Above the freezing temperature $T > T_{\rm f}$, $\Gamma_{\rm qel}$ exhibits monotonously increasing behavior as the temperature is elevated.
This is normal behavior attributable to increased thermal fluctuations.
It, however, does not obey the Korringa law $\Gamma_{\rm qel} \propto T$, which is expected for fluctuations of localized moments induced through the $s-d$ interactions in the normal metal~\cite{abr61}.
Instead, the temperature dependence for $T > T_{\rm f}$ could be expressed by the Arrhenius law biased by the zero temperature relaxation rate $\Gamma_0$:
\begin{equation}
\Gamma_{\rm qel} = \Gamma_0 + \Gamma_1 \exp(-E_{\rm A}/k_{\rm B}T).
\end{equation}
The solid line in the Fig.~7 shows the best fit of the observed widths to the above equation; optimum values for the parameters are estimated as $E_{\rm A} = 3.7(6)$~meV, $\Gamma_0 = 0.46(5)$~meV, and $\Gamma_1 = 1.7(2)$~meV.
This Arrhenius-type behavior of the quasielastic peak width has been occasionally observed in the metallic spin glasses~\cite{mot88,mot91}, and interpreted as that energy barriers with the height $E_{\rm A}$ are responsible for the glassy (freezing) behavior.

Although the temperature dependence of the relaxation rate for $T > T_{\rm f}$ is well described by the Arrhenius behavior, there is a distinct and prominent feature that differentiates the present Zn-Fe-Sc quasicrystal from the metallic spin glasses; $\Gamma_{\rm qel}$ stays considerably large even below $T_{\rm f}$, which is in striking contrast to the infinitely small $\Gamma$ at $T_{\rm f}$ in metallic spin glasses.
The dashed line in the figure shows the estimate of the thermal energy $k_{\rm B}T$.
Note that $\Gamma_{\rm qel}$ is larger than the thermal energy at low temperatures, suggesting that the spin fluctuations may not be due to thermal origin but may be certain quantum fluctuations.
In view of the single-site nature of the spin fluctuations, this may most likely be due to on-site scattering of localized moments by conduction electrons, and thus suggest significant interaction between the localized $3d$ moments and conduction electrons.

\begin{figure}
\includegraphics[scale=0.32, angle=-90]{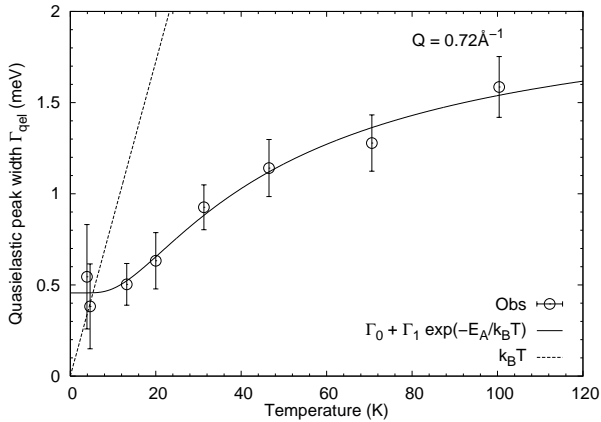}
\caption{Temperature dependence of the quasielastic peak width $\Gamma_{\rm qel}$ obtained by the resolution-convolution fitting.
The solid line stands for a fit to $\Gamma_0 + \Gamma_1 \exp(-E_{\rm A}/k_{\rm B}T)$, whereas the dashed line denotes the thermal energy $k_{\rm B}T$. }
\end{figure}

\section{Conclusions}
Spin correlations and fluctuations have been studied in the newly discovered $3d$-transition-metal based quasicrystal Zn-Fe-Sc.
The magnetic diffuse scattering has been observed at low temperatures.
A quantitative comparison shows that the strength of the inter-site spin correlations in Zn-Fe-Sc is similar in strength to that in the rare-earth based magnetic quasicrystals.
In the inelastic scattering experiment, the $Q$-independent quasielastic signal has been observed in the entire temperature range of the present study, indicating intrinsic single-site relaxational fluctuations.
The spin relaxation rate $\Gamma_{\rm qel}$ at higher temperatures $T > T_{\rm f}$ shows Arrhenius-type behavior, which is an occasionally observed feature in the metallic spin glasses.
Nevertheless, it shows the finite width ($\simeq 0.46$~meV) below the macroscopic freezing temperature $T_{\rm f}$, in contrast to the infinitesimally small $\Gamma_{\rm qel}$ for metallic spin glasses.
This finite $\Gamma_{\rm qel}$ suggests remaining quantum fluctuations due to the scattering of the $3d$ moments by the conduction electrons.

\begin{acknowledgments}
The present authors thank Dr. A. P. Tsai for stimulating discussion.
This work was partly supported by a Grant-in-Aid for Encouragement of Young Scientists (B) (No. 16760537) and by a Grant-in-Aid for Creative Scientific Research (No. 16GS0417) from the Ministry of Education, Culture, Sports, Science and Technology of Japan.
\end{acknowledgments}


\end{document}